\documentclass[sigconf]{acmart}
\usepackage{multicol, multirow}
\usepackage{subfigure}
\usepackage{algorithm}
\usepackage{algorithmic}
\usepackage{color}
\usepackage{hyperref}
\usepackage{url}
\usepackage[utf8]{inputenc}
\AtBeginDocument{%
  }

\copyrightyear{2025}
\acmYear{2025}
\setcopyright{acmlicensed}\acmConference[WWW Companion '25]{Companion
Proceedings of the ACM Web Conference 2025}{April 28-May 2, 2025}{Sydney,
NSW, Australia}
\acmBooktitle{Companion Proceedings of the ACM Web Conference 2025 (WWW
Companion '25), April 28-May 2, 2025, Sydney, NSW, Australia}
\acmDOI{10.1145/3701716.3715206}
\acmISBN{979-8-4007-1331-6/2025/04}

\begin{document}

%%
%% The "title" command has an optional parameter,
%% allowing the author to define a "short title" to be used in page headers.
\title{A Contextual-Aware Position Encoding for Sequential Recommendation}

%%
%% The "author" command and its associated commands are used to define
%% the authors and their affiliations.
%% Of note is the shared affiliation of the first two authors, and the
%% "authornote" and "authornotemark" commands
%% used to denote shared contribution to the research.
\author{Jun Yuan}
\authornote{Both authors contributed equally to this research.}
\email{yuanjun25@huawei.com}
\affiliation{%
  \institution{Huawei Technologies Co., Ltd.}
  \city{Shenzhen}
  \state{Guangdong}
  \country{China}
}

\author{Guohao Cai}
\authornotemark[1]
\email{caiguohao1@huawei.com}
\affiliation{%
  \institution{Huawei Noah’s Ark Lab}
  \city{Shenzhen}
  \state{Guangdong}
  \country{China}
}

\author{Zhenhua Dong}
\email{dongzhenhua@huawei.com}
\affiliation{%
  \institution{Huawei Noah’s Ark Lab}
  \city{Shenzhen}
  \state{Guangdong}
  \country{China}
}

%%
%% By default, the full list of authors will be used in the page
%% headers. Often, this list is too long, and will overlap
%% other information printed in the page headers. This command allows
%% the author to define a more concise list
%% of authors' names for this purpose.
\renewcommand{\shortauthors}{Jun Yuan, Guohao Cai, and Zhenhua Dong}
%%
%% The abstract is a short summary of the work to be presented in the
%% article.
\begin{abstract}
Sequential recommendation (SR), which encodes user activity to predict the next action, has emerged as a widely adopted strategy in developing commercial personalized recommendation systems. A critical component of modern SR models is the attention mechanism, which synthesizes users' historical activities. This mechanism is typically order-invariant and generally relies on position encoding (PE). Conventional SR models simply assign a learnable vector to each position, resulting in only modest gains compared to traditional recommendation models. Moreover, limited research has been conducted on position encoding tailored for sequential recommendation, leaving a significant gap in addressing its unique requirements. To bridge this gap, we propose a novel \underline{C}ontextual-\underline{A}ware \underline{P}osition \underline{E}ncoding method for sequential recommendation, abbreviated as \textbf{CAPE}. To the best of our knowledge, CAPE is the first PE method specifically designed for sequential recommendation. Comprehensive experiments conducted on benchmark SR datasets demonstrate that CAPE consistently enhances multiple mainstream backbone models and achieves state-of-the-art performance, across small and large scale model size. Furthermore, we deployed CAPE in an industrial setting on a real-world commercial platform, clearly showcasing the effectiveness of our approach. Our source code is available at \url{https://github.com/yjdy/CAPE}.
\end{abstract}

%but two critical questions remain under-explored: firstly the real value of incorporating position encoding in SR models; secondly the necessity of proposing specific PE method for SR.

%%
%% The code below is generated by the tool at http://dl.acm.org/ccs.cfm.
%% Please copy and paste the code instead of the example below.
%%
%%
%% Keywords. The author(s) should pick words that accurately describe
%% the work being presented. Separate the keywords with commas.
%% A "teaser" image appears between the author and affiliation
%% information and the body of the document, and typically spans the
%% page.
%%
%% This command processes the author and affiliation and title
%% information and builds the first part of the formatted document.
\maketitle

\section{Introduction}
Recommendation systems play a pivotal role in current online content platforms and e-commerce. The recommendation algorithm is a complex problem that requires extracting user interests to predict future behaviors across various items. Nowadays, sequential recommendation (SR) has occupied a dominant position in commercial recommendations systems, including e-commerce~\cite{chen2019behaviorsequencetransformerecommerce}, social media~\cite{xia2023transacttransformerbasedrealtimeuser}, news/video feeds~\cite{10.1145/3477495.3531862}, and online advertising~\cite{10.1145/3219819.3219823}. The goal of sequential recommendation systems is to combine personalized models of user behavior (based on historical
activities) with a notion of "context" derived from users’ recent actions~\cite{kang2018selfattentivesequentialrecommendation}. Sequential recommendation has been explored for years and various SR models have been proposed~\cite{10.1145/2766462.2767694,bert4rec,10.1609/aaai.v33i01.33015941}. 

In recent years, the attention mechanism has emerged as a critical component of sequential recommendation models, enabling interactions among items in a sequence despite being inherently order-invariant. Incorporating position encoding (PE) addresses this limitation by enabling position-aware item representations~\cite{dufter2021positioninformationtransformersoverview}. Various methods have been developed to incorporate position information into attention in natural language processing (NLP). However, the mainstream PE approach in SR primarily involves assigning a learnable vector to each position~\cite{chen2019behaviorsequencetransformerecommerce,tin}. Moreover, some widely-used sequential recommendation models, such as DIN~\cite{10.1145/3219819.3219823} and DIEN~\cite{10.1609/aaai.v33i01.33015941}, do not employ any PE methods at all. As a result, these attempts have only achieved modest improvements over traditional recommendation models.

% But two critical questions remain
% under-explored: firstly the real value of incorporating position en-
% coding in SR models; secondly the necessity of proposing specific
% PE method for SR.

\begin{figure}[tp]
    \centering
    \includegraphics[width=0.45\textwidth,height=0.3\textwidth]{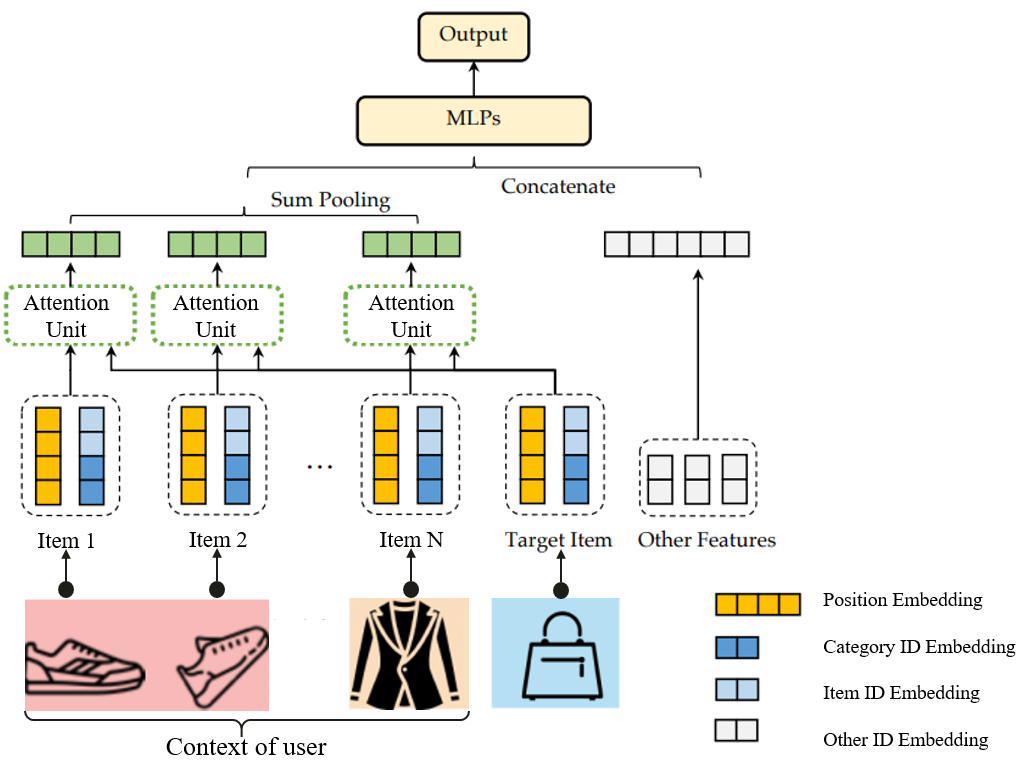}\vspace{-2mm}
    \caption{Overview of mainstream sequential recommendation models. We only demonstrate the target attention mechanism, which can be seen as calculating the attention output of last item in self-attention}\label{fig:sr_overview}
    \vspace{-4mm}
\end{figure}

Simply adopting PE methods from NLP may not be suitable for sequential recommendation due to three key differences between models in SR and those in NLP: 

\textbf{First}, as illustrated in Figure~\ref{fig:sr_overview}, item embeddings in SR are generally constructed by concatenating heterogeneous feature embeddings, such as item ID and category ID embeddings, whereas token embeddings in language models reside in a unified hidden space. It is necessary to fuse heterogeneous features when adding position information. \textbf{Second}, while dot product attention is the predominant method in language models, SR employs a variety of attention mechanisms, including DIN attention~\cite{10.1145/3219819.3219823} and bilinear attention~\cite{10.1609/aaai.v33i01.33015941} etc. \textbf{Finally}, representing various levels of position information, such as session-level and time-span-level, is crucial in SR to capture underlying user intents, extending beyond decoding relationships between items. For instance, users may occasionally concentrate on a particular subset of items within a session or time period, driven by a strong underlying intent. However, this poses a challenge for current PE methods, which assign an item-independent embedding vector to each position and simply add it to the corresponding item representations~\cite{dufter2021positioninformationtransformersoverview}.

These three fundamental differences between SR and natural language declare the necessity of designing specialized PE methods tailed to sequential recommendation models. Moreover, two unique characteristics of SR should be taken into account deeply during the design process. The first one is that the absolute order of items may not as important as token order in natural language. For example, varying the order of consecutively purchased items by the same intent often result in identical product recommendations. The second one is that online recommendation systems often demand high efficiency from models in both training and inference.

As discussed above, directly adopting position encoding methods from NLP is unlikely to significantly enhance SR models. Unfortunately, such issue has received minimal attention in the field of sequential recommendation research. To fill this gap, we propose a contextual-aware position encoding, denoted as \textbf{CAPE}, specifically designed for SR. CAPE dynamically assigns position encoding based on user history, making it a context-dependent and efficient approach tailored to the unique needs of sequential recommendation. Our contributions in this paper are summarized as follows:

% In this way, CAPE can represent various levels of position abstraction at the same time like CoPE~\cite{golovneva2024contextualpositionencodinglearning}. 

\begin{itemize}
    \item We propose CAPE, a Contextual-Aware Position Encoding method tailored for sequential recommendation. To the best of our knowledge, it is the first approach specifically designed for sequential recommendation models. CAPE efficiently captures and represents multiple levels of positional abstraction within the user context, enhancing the model's ability to leverage sequential information.
    \item We introduce a dissimilarity measure to assign positions to context items, accounting for the unique characteristics of SR. Additionally, we design a gate architecture combined with interpolation to efficiently fuse heterogeneous embeddings, thereby enhancing model performance and computational efficiency simultaneously.
    \item Extensive experiments on public benchmark sequential recommendation datasets consistently demonstrate CAPE's ability to improves various SR backbones, including small and large scale recommendation models. Furthermore, we deployed CAPE on a commercial recommendation system, and online A/B testing results validated the effectiveness of our proposed method.
\end{itemize} 

\begin{figure*}[tp]
    \centering
    \includegraphics[width=0.9\textwidth,height=0.25\textwidth]{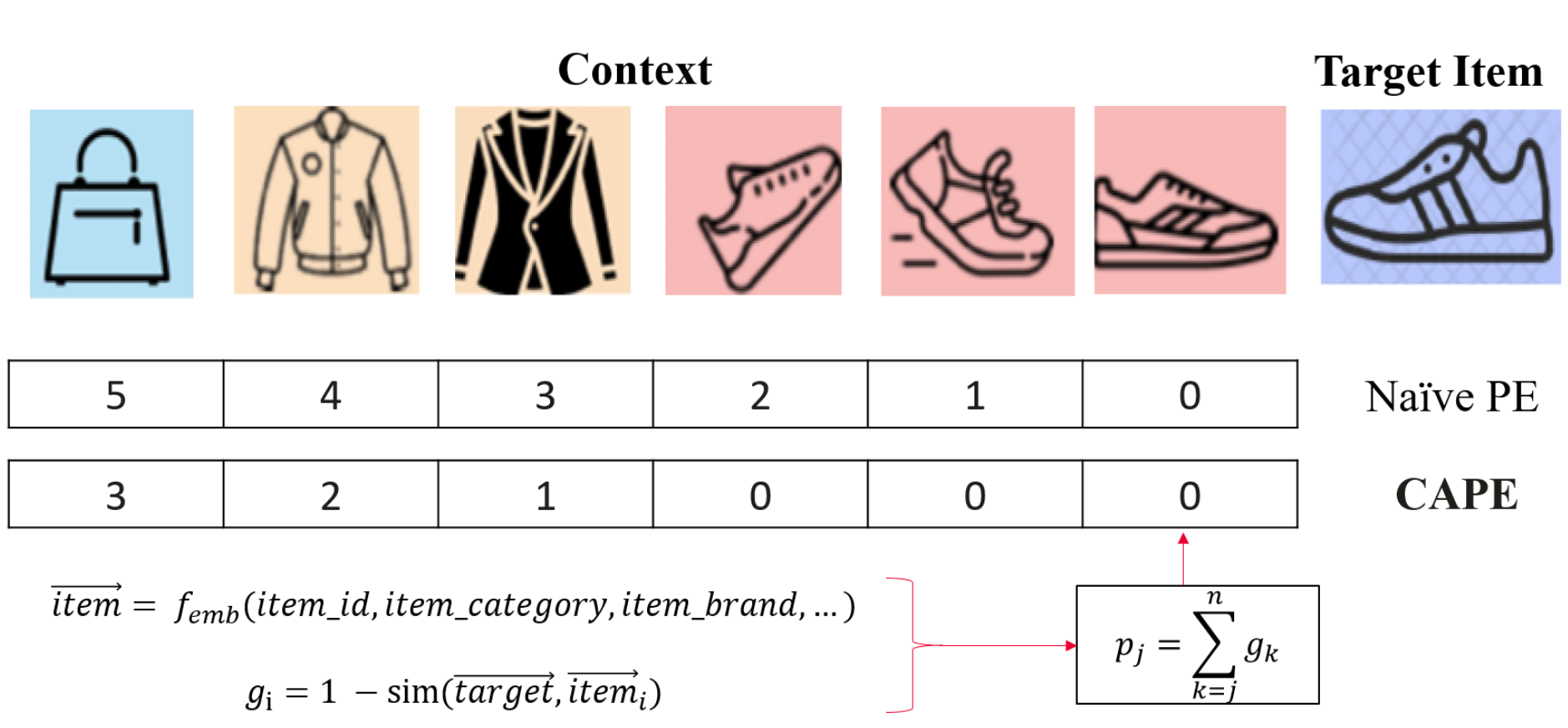}\vspace{-3mm}
    \caption{Overview of CAPE. CAPE first computes the dissimilarity of target item and context items, and then accumulating the dissimilarity values to get the position of each context item. In this way, CAPE tends to assign same position to similar context items and vice versa. CAPE can be adopted to fit self-attention easily.}\label{fig:cape_overview}
    \vspace{-3mm}
\end{figure*}

\section{Related Work}

\subsection{Sequential Recommendation} 
Early sequential recommendation are based on Markov chains~\cite{10.1145/2766462.2767694}, RNN/CNN~\cite{hidasi2016sessionbasedrecommendationsrecurrentneural,10.1145/3289600.3290975}. Following the significant advancements in NLP, various attention-based SR models have been proposed~\cite{10.1145/3219819.3219823,10.1609/aaai.v33i01.33015941,kang2018selfattentivesequentialrecommendation,chen2019behaviorsequencetransformerecommerce,10.1145/3340531.3411954,bert4rec}, which have become a critical component of modern commercial recommendation systems. Attention-based SR models can be categorized into two categories: target attention based models~\cite{10.1145/3219819.3219823,10.1609/aaai.v33i01.33015941} and self-attention based models~\cite{kang2018selfattentivesequentialrecommendation,chen2019behaviorsequencetransformerecommerce,10.1145/3340531.3411954,bert4rec,Liu2021NoninvasiveSF}. Position encoding is typically considered only in attention-based models, as the attention mechanism is \textit{order-invariant}.

\textbf{Target attention based models}. Deep interest network (DIN)~\cite{10.1145/3219819.3219823} utilizes an attention mechanism to locally activate historical behaviors with respect to the given target item, capturing the diversity of user context to form the representation of target item. DIEN~\cite{10.1609/aaai.v33i01.33015941} introduces the interest evolving layer, employing a GRU with an attentional update gate (AUGRU) to model the interest evolving process relative to the target item. TIN~\cite{tin} incorporates  target-aware temporal encoding, target-aware attention, and target-aware representation to capture semantic-temporal correlation between user behaviors and the target item. While DIN and DIEN do not consider position information in context, TIN employs an absolute PE methods, assigning each position a learnable embedding vector.

\textbf{Self-attention based models}. Due to the great achievements of the \textit{Transformer} architecture in NLP and CV, plenty of researches have sought to apply self-attention mechanisms to sequential recommendation problems. SASRec~\cite{kang2018selfattentivesequentialrecommendation} utilizes multi-head \textit{Transformer} decoder to generate next item. Bert4Rec~\cite{bert4rec} uses a Cloze objective loss for sequential recommendation through the bidirectional self-attention mechanism. BST~\cite{chen2019behaviorsequencetransformerecommerce} use the Transformer model to capture the sequential patterns underlying users’ behavior sequences for recommendation. DMIN~\cite{10.1145/3340531.3412092} combines both target and self attention, leveraging a transformer architecture to model context sequence while using a DIN to integrate target embeddings with context sequence embeddings. DMIN can be viewed as an update version of DIEN. ALl self-attention based models use PE methods, although they generally employ a na\"ive absolute PE approach. 

\textbf{Scaling law in Recommendation}. In recent years, the scaling of model size and data scale has been a vital point in various fields, e.g., NLP, and CV etc. some approaches~\cite{10.1145/3340531.3411954,10.1145/3580305.3599519,10.1145/3581783.3611967} incorporate item contexts such as item text, visual information or categorical features into ID embeddings to improve data scale by utilizing LLMs. S3-Rec~\cite{10.1145/3340531.3411954} pre-trains sequential models with mutual information maximization to capture the correlations among attributes, items, subsequences, and sequences. MISSRec~\cite{10.1145/3581783.3611967} proposes a novel multi-modal pre-training and transfer learning framework, effectively addressing the cold-start problem and enabling efficient domain adaptation. In this paper, we focus solely on the ID features, though CAPE can be easily extended to multi-modal features. 

There are also some efforts to valid model size in RecSys area (e.g., HSTU~\cite{zhai2024hstu}, HLLM~\cite{chen2024hllmenhancingsequentialrecommendations}, MARM~\cite{lv2024marmunlockingfuturerecommendation}). The recent RecSys Scaling-laws studies are greatly changing the learning paradigms of conventional recommendation models. Current researches indicate that complex and deeper networks have potential to dominate future recommendation systems. We also evaluate CAPE with different model size in this paper.

\subsection{Position Encoding} PE emerged as an important research topic with the rise of the Transformer architecture. The original Transformer paper by~\cite{10.5555/3295222.3295349} introduced absolute PE, using fixed vectors generated through sinusoidal functions at varying frequencies. Early large language models (LLMs)~\cite{Radford2018ImprovingLU} also adopted absolute PE, enhancing it by adding learnable embedding vectors corresponding to each relative position within hidden representations. This approach remains the most widely used PE method in current SR models.

The relative position encoding method was later introduced in~\cite{Shaw_Uszkoreit_Vaswani_2018}. RoPE~\cite{10.1016/j.neucom.2023.127063} extends this concept by rotating query and key vectors by angles proportional to their absolute positions, making attention logits dependent on positional differences. These PE methods are context-independent, \emph{i.e.}, they assign positions based solely on the order of items in a sequence. In contrast, CoPE~\cite{golovneva2024contextualpositionencodinglearning} adopts a context-dependent approach by measuring positions relative to the context, deviating from the traditional token-based position paradigm. CAPE is also a relative and context-dependent PE method. However, instead of relying on similarity as in CoPE, it uses dissimilarity to calculate positions along with a fusion method, tailored specifically to the characteristics of SR. 

Current researches use position encoding with little attention to them. Lopez-Avila et al~\cite{lopezavila2024positionalencodingcontextstudy} evaluated the effect of choosing different methods and found that PE can be a powerful tool to improve recommendation performance. However, as discussed in Introduction, directly adopting PE from NLP may not be suitable for SR, and designing PE specifically for SR is under-explored nowadays.

\section{Problem Description}

\subsection{Preliminaries}
The attention outputs $\mathbf{o}_i$ for each $i$-th item in SR are generally expressed as follows:
$$\mathbf{o}_i = \sum_{j}^{i} a_{ij}\mathbf{h}_{j}$$ where $a_{ij}$ is the attention weight between the \emph{i}-th and \emph{j}-th items. Since the attention mechanism is order-invariant, it is essential to incorporate a position encoding mechanism~\cite{10.5555/2969442.2969512}. PE methods can be broadly categorized into two groups: absolute and relative PE. 

\textbf{Absolute PE} simply adds a vector representing an absolute position \emph{j} to the hidden states, \emph{i.e.} $\mathbf{h}_{j} \leftarrow \mathbf{h}_{j} + P(j)$. The function $P(j)$ can be implemented through an embedding layer that assigns a unique learnable vector $\mathbf{e}[j]$. We refer to this method as Na\"ive PE in this paper. Currently, Na\"ive PE is the most widely used method in SR, according to our observations. %Alternatively, PE can be implemented as a fixed mapping using sinusoidal functions with varying frequencies~\cite{10.5555/3295222.3295349}.

\textbf{Relative PE}~\cite{Shaw_Uszkoreit_Vaswani_2018} depends on both the position of the current token \emph{i} and the position \emph{j} of the token being attended to. Therefore, it must be implemented within the attention layer, such as: $$a_{ij} = Softmax(\mathbf{q}_{i}^{\top}(\mathbf{k}_{j}+P(i-j)), j\in [1,i]$$ $\mathbf{q}$ is the query vector in attention mechanism, and $\mathbf{k}$ is key vector. $P(i-j)$ means the relative position between \emph{i}-th token and \emph{j}-th token. Relative PE is especially suited for processing unbounded sequences. This superiority can be very important in SR, since length of context can be much longer than sentence.

It should be noted that item embeddings in SR are generally constructed by concatenating heterogeneous feature embeddings, whereas token embeddings in language models reside in a unified hidden space. It is necessary to fuse heterogeneous features when adding position information in user context.

\subsection{Problem Definition}
We focus on the task of sequential recommendation, which is formally defined as follows: Given a user $u \in \mathcal{U}$ and a sequence of the user's historical interactions (referred to as the context) $U = \{I_1, I_2, \dots, I_n\}$ arranged in chronological order, the objective is to predict the next item $I_{n+1}$, where $n$ is the length of $U$. $\mathcal{I}$ represents the set of all items, \emph{i.e.} $\forall I \in \mathcal{I}$. 

Each item $I$ is associated with an ID and additional features (e.g., category, tags, etc.). The hidden representation of the user context is denoted as $\{\mathbf{h}_{1}, \dots, \mathbf{h}_{n}\}$, capturing the sequential patterns and interactions of the user.

\begin{table}[htp]
\caption{Summary of notations}\vspace{-4mm}
\label{tab:notations}
\centering
\begin{tabular}{c|c}
\toprule
\textbf{Notation}&\textbf{Description}\\\hline
$I$, $t$ &Context item and target item\\
$\mathbf{h}$, $\mathbf{t}$ &Embedding of context item and target item\\
% $t$&Current training step\\
$n$&Length of context\\
$p_{j}$&Position of \emph{j}-th context item\\
% $\mathbf{o}$&Output of attention unit\\
$sim_{item}$&Similarity function between items\\
$sim_{pos}$&Similarity function between item and position\\
\bottomrule
\end{tabular}
\vspace{-4mm}
\end{table}

\section{Context-Aware Position Encoding}\label{method}
%In this section, we provide a detailed explanation of CAPE, focusing on how positions and their corresponding embeddings are assigned to different items. 

As discussed in the introduction, we believe that sequential recommendation requires a position encoding that addresses two key challenges: 1) Identifying which context items are more relevant for the target item, and 2) Effectively and efficiently fusing item and position embeddings. In the following, we will introduce the solutions to the two problems. For simplicity, we will focus on the formulation of target attention, which can be easily extended to self-attention. The overview of CAPE is demonstrated in Figure~\ref{fig:cape_overview}.

\subsection{Context-aware Dissimilarity}
% 可能的故事，item embedding和position embedding 不在一个空间，所以出于简单考虑，这里只考虑dot production

First, in our method, the positions are determined in a context-dependent manner, rather than being a simple embedding vector assigned to each position. Our approach operates it by first determining which tokens should be considered when measuring distance using their embedding vectors. Specifically, a gate value, to represents dissimilarity, is computed for target item embedding $\mathbf{t}$ and every context item embedding $\mathbf{h}_{j}$
\begin{equation}
    g_{j} = 1-\sigma(sim(\mathbf{t}, \mathbf{h}_{j})), j \in [1, n]
\end{equation} 
where $n$ denotes the length of context, $sim$ is the similar function, and $\sigma$ is the sigmoid function, $\mathbf{t}$ is the hidden representation of target item, and $\mathbf{h}_{j}$ represents the embedding of context item at position \emph{j}. A gate value of 1 means that the key will be counted in the position measurement, while a gate value of 0 implies it will be disregarded.

Next, we compute position values by summing the gate values between the current and the target item
\begin{equation}
    p_{j} = \sum_{k=j}^{n} g_{k}
\end{equation} Note that if gates are always 1, then $p_{j} = j+1$ and we recover item-based relative positions. Thus, our method can be viewed as a generalization of relative PE, indicating CAPE has potential to processing unbounded sequences~\cite{Shaw_Uszkoreit_Vaswani_2018}. As displayed in Figure~\ref{fig:cape_overview}, related context items intent to be assigned same position, while irrelevant context items are likely to be assigned distinct positions. In this manner, CAPE enables SR models to attend to varying levels of position abstraction simultaneously.

Unlike conventional PE, our position values $p_{j}$ are not restricted to integers and can take fractional values due to the sigmoid function. Consequently, we cannot use an embedding layer to convert a position value to a vector like in the relative PE. Inspired by CoPE~\cite{golovneva2024contextualpositionencodinglearning}, we use interpolation between integer values. Initially, we assign a learnable embedding vector $\mathbf{e}[p]$ to each integer position $p \in [1, n]$. The embedding for position $p_{j}$ is then computed as a simple interpolation between the two nearest integer embeddings:
\begin{equation}
    \mathbf{e}[p_{j}] = (p_{j} - \lfloor p_{j} \rfloor)\mathbf{e}[\lceil p_{j} \rceil] + (1 - p_{j} + \lfloor p_{j} \rfloor)\mathbf{e}[\lfloor p_{j} \rfloor]
\end{equation} In this way, we can assign position embedding to each context item. We will discuss how to fuse item and position embedding in the following subsection.

\subsection{Fusing item and position embedding}

In the field of natural language processing, attention logits are typically computed by simply adding position embeddings and sequence embeddings, like $a_{j} = Softmax(sim(\mathbf{t}, (\mathbf{h}_{j} + \mathbf{e}[p_{j}])))$. Conventional SR models simply follow the trick, \emph{i.e.} straightforwardly adding position and item embedding. 

However, the implicit assumption behind this trick is that both types of embeddings lie in the same hidden space. But, as mentioned in introduction, this assumption may not hold in SR, because item embedding in SR generally constructed by concatenating heterogeneous embedding, such as \textit{item id}, \textit{categories} and \textit{tags} etc. Item embedding and position embedding may reside in different hidden spaces. Furthermore, unlike \emph{Transformers}, where attention is computed solely through dot production, multiple types of attention have played important role in sequential recommendation.

Therefore, fusing heterogeneous embeddings and transferring position embeddings into the space of context item embeddings is essential. We consider a more general formulation as follows:
\begin{equation}
    a_{j} = Softmax(sim_{item}(\mathbf{t},f(\mathbf{h}_{j},\mathbf{e}[p_{j}])))
\end{equation} $sim_{item}$ refers to the similarity function between the target item and the context item, which can involve various attention mechanisms in SR, such as DIN attention and Bilinear attention etc, depending on the specific use case. To achieve efficient fusion of item and position embedding, we employ a gating architecture and interpolation for the function $f(\cdot)$. 

\textbf{Gate Architecture}. To fuse heterogeneous features and align the embeddings of item and position into the same hidden space, we adopt a gating architecture, similar to that in \cite{yuan19transgate}, to reduce dimension of target embedding and extract relevant information at the same time.
\begin{equation}
    \mathbf{t}' = \mathrm{SiLU}(\mathbf{W}\mathbf{t} + \mathbf{b})
\end{equation} $\mathbf{W} \in \mathbb{R}^{d\times d_{pos}}$ and $\mathbf{b} \in \mathbb{R}^{d_{pos}}$. $d$ is the dimension of item embedding and $d_{pos}$ represents the dimension of position embedding. Naturally, we can simply set $d_{pos} < d$ to reduce computational and memory costs, according to the actual situation. We only use a linear layer to fuse features, because of the large embedding size and high efficiency demand of online recommendation systems. 

\textbf{Interpolation}. Computing and storing vectors $\mathbf{e}[p_{j}]$ still need extra computation and memory. To optimize this, we first compute the $sim_{pos}(\mathbf{t}', \mathbf{e}[p])$ for all position $p$:
\begin{equation}
    z[p] = sim_{pos}(\mathbf{t}', \mathbf{e}[p_{j}]) \ for\ p \in [1, n]
\end{equation} In this paper, we use dot production as the $sim_{item}$, and we will explore other type of attention in future work. Then, we further enhance computational efficiency by interpolation:
\begin{equation}
    z[p_{j}] = (p_{j} - \lfloor p_{j} \rfloor)z[\lceil p_{j} \rceil] + (1 - p_{j} + \lfloor p_{j} \rfloor)z[\lfloor p_{j} \rfloor]
\end{equation} $z[p_{j}]$ is a scalar and needs to be added to the attention logits of \emph{j}-th item in context.

Finally, the attention weights on the context item embedding $\mathbf{h}_{j}$ are computed as follows:
\begin{equation}
    a_{j} = \frac{\exp(sim_{item}(\mathbf{t},\mathbf{h}_{j}) + z[p_{j}])}{\sum_{i=0}^{n}\exp(sim_{item}(\mathbf{t},\mathbf{h}_{i}) + z[p_{i}])}
\end{equation} With the attention weights, a sum pooling is then adopted to aggregate the user’s history: $$\mathbf{o} = \sum_{j}^{n} a_{j}\mathbf{h}_{j}$$

\section{Experiment}

\subsection{Experiment Setups}
\subsubsection{Baseline PE method} We compare CAPE with the following mainstream PE methods in the field of sequential recommendation. We re-implement following PE methods and integrate them with SOTA SR backbones.

\begin{itemize}
    \item \textbf{None} means no position encoding method in SR models. If a backbone originally contains a PE method, we will remove the related code.
    \item \textbf{Na\"ive PE} is implemented by an embedding layer that assigns a unique learnable vector to each position. The dimension of position embedding is same as item embedding. We add item and position embedding before network. 
    \item \textbf{CoPE}~\cite{golovneva2024contextualpositionencodinglearning}. We re-implement CoPE according to the paper, and change the attention code of backbones to integrate this PE method. The dimension of position embedding is same as item embedding too.
    \item \textbf{RoPE}~\cite{10.1016/j.neucom.2023.127063} rotates context item vectors in target attention and the concatenation of context and  target item in self-attention. In all experiments, we set the maximum length of RoPE same as the maximum length of context.
\end{itemize}

\subsubsection{SR Backbones}
In our experiments, we integrate above PE methods with the following mainstream SR backbones. We use the code released from FuxiCTR~\cite{DBLP:conf/cikm/ZhuLYZH21}\footnote{https://github.com/reczoo/FuxiCTR} and change the position encoding part of each backbones.
\begin{itemize}
    \item \textbf{DIN}~\cite{10.1145/3219819.3219823} proposed a new attention mechanism (called din attention in this paper), and this mechanism combine embedding of target item, embedding of context, element-wise production. DIN do not use PE method, and we add PE methods into the attention unit.
    \item \textbf{DIEN}~\cite{10.1609/aaai.v33i01.33015941} uses a RNN to encode item sequence, then use a target attention unit to get attention output. DIEN use three attention method: dot production, bilinear and din attention. DIEN do not use PE method, and we add PE method into attention unit in experiment.
    \item \textbf{BST}~\cite{chen2019behaviorsequencetransformerecommerce} uses Transformer to capture deeper relationship of input features. And it adds position as an input feature of each item in the bottom layer before it is projected as a low-dimensional vector.
    \item \textbf{DMIN}~\cite{10.1145/3340531.3412092} tries to combine both target and self attention. It uses a self attention to model context sequence and uses a DIN to integrate target item and context sequence. DMIN adds position as an input feature of each item too. We add PE in both target attention and self attention with same position embedding dimension.
    \item \textbf{SASRec}~\cite{kang2018selfattentivesequentialrecommendation}  is a self-attention based sequential recommendation model, which uses the multi-head attention mechanism to recommend the next item.
\end{itemize}

\begin{table*}[htp]
    \centering
    \caption{Evaluation in Public Benchmark Datasets, AmazonElectronics and KuaiVideo. AUC, gAUC and logloss values are averages over 5 random seeds and only keep 6 decimal. The best results within the same backbone are bold with $p<0.05$, and the underline indicates the second best.}
    % \vspace{-2mm}
    \resizebox{0.76\textwidth}{!}{\begin{tabular}{c|c|ccc|ccc}
    \toprule
         \multirow{2}{*}{\textbf{SR Model}}&\multirow{2}{*}{\textbf{PE}}&\multicolumn{3}{c|}{AmazonElectronics}&\multicolumn{3}{c}{KuaiVideo}  \\
         \cmidrule(r){3-5} \cmidrule(r){6-8}
         &&gAUC$\uparrow$&AUC$\uparrow$&logloss$\downarrow$&gAUC$\uparrow$&AUC$\uparrow$&logloss$\downarrow$\\\hline
         % \multirow{4}{*}{DIN}&&&&&&\\
         \multirow{5}{*}{DIN}&None&0.883526&0.886028&0.43019&0.661646&0.741604&0.447621\\
         &Na\"ive&0.883947&0.886702&\underline{0.428511}&0.661123&0.743029&0.441684\\
         &RoPE&0.884544&\underline{0.8872200}&0.429256&\underline{0.664594}&\underline{0.745957}&\underline{0.439012}\\
         &CoPE&\underline{0.884549}&0.886706&0.430099&0.661646&0.742936&0.442531\\
         &CAPE&\textbf{0.885698}&\textbf{0.888156}&\textbf{0.428468}&\textbf{0.665215}&\textbf{0.745973}&\textbf{0.438510}\\\hline
         % DIEN&&&&&&\\
         \multirow{5}{*}{DIEN}&None&0.884014&0.886774&0.428798&0.661032&0.743491&0.437697\\
         &Na\"ive&0.885397&0.887903&\underline{0.426713}&0.659564&\underline{0.744395}&\underline{0.435241}\\
         &RoPE&\underline{0.887128}&\underline{0.889513}&0.426858&0.661536&0.744089&0.438067\\
         &CoPE&0.885736&0.888723&0.426941&\underline{0.661589}&0.744392&0.435911\\
         &CAPE&\textbf{0.887736}&\textbf{0.889723}&\textbf{0.425941}&\textbf{0.662178}&\textbf{0.744486}&\textbf{0.434926}\\\hline
         % BST&&&&&&\\
         \multirow{5}{*}{BST}&None&0.878645&0.879191&0.464533&0.661409&0.741465&0.446131\\
         &Na\"ive&0.881701&0.884188&0.430641&0.660665&0.744091&0.435769\\
         &RoPE&\underline{0.882918}&\underline{0.885657}&\textbf{0.429100}&\underline{0.662909}&\underline{0.745502}&\textbf{0.432553}\\
         &CoPE&0.882166&0.884399&\underline{0.430178}&0.660777&0.744202&0.435414\\
         &CAPE&\textbf{0.883349}&\textbf{0.886499}&0.430582&\textbf{0.664139}&\textbf{0.746326}&\underline{0.433429}\\\hline
         % SASRec&&&&&&\\
         \multirow{5}{*}{SASRec}&None&0.879654&0.879692&0.460631&0.659623&0.744130&0.437802\\
         &Na\"ive&0.879493&0.882471&0.434959&0.661535&0.743983&0.436454\\
         &RoPE&\underline{0.882185}&\underline{0.884663}&0.433952&\underline{0.661662}&\underline{0.745082}&\underline{0.434806}\\
         &CoPE&0.880922&0.883757&\underline{0.433889}&0.654668&0.741277&\textbf{0.433345}\\
         &CAPE&\textbf{0.882793}&\textbf{0.885344}&\textbf{0.431368}&\textbf{0.663006}&\textbf{0.745512}&0.435691\\\hline
         % DMIN&&&&&&\\
         \multirow{5}{*}{DMIN}&None&0.885558&0.887029&0.427322&0.659194&0.743807&0.435644\\
         &Na\"ive&0.883131&0.885329&0.431241&\underline{0.661405}&\underline{0.745304}&\textbf{0.434112}\\
         &RoPE&0.884383&0.886762&\underline{0.425995}&0.660126&0.745036&0.433466\\
         &CoPE&\underline{0.885583}&\underline{0.887662}&0.426295&0.659226&0.744025&0.434486\\
         &CAPE&\textbf{0.885703}&\textbf{0.888053}&\textbf{0.424144}&\textbf{0.662567}&\textbf{0.746088}&\underline{0.434272}\\
    \bottomrule
    \end{tabular}}
    \vspace{-2mm}
    \label{tab:main_result}
\end{table*}

\subsubsection{Dataset}
To evaluate the effeciveness of CAPE, we conduct plenty of experiments in three open benchmark user behavior datasets provide by reczoo\footnote{https://huggingface.co/reczoo}: AmazonElectronics, KuaiVideo and AmazonBooks.
\begin{itemize}
    \item \textbf{AmazonElectronics} is a subset of the Amazon dataset~\cite{10.1145/2872427.2883037}, a widely used benchmark dataset for recommendation. We follow the DIN~\cite{10.1145/3219819.3219823} to preprocess the dataset. Specifically, the AmazonElectronics contains 1,689,188 samples, 192,403 users, 63,001 goods and 801 categories. Features include goods\_id, category\_id,and their corresponding user-reviewed sequences: goods\_id\_list, category\_id\_list.
    \item \textbf{KuaiVideo} is another open dataset for short video recommendation. We follow the work~\cite{10.1145/3343031.3350950} to obtain the preprocessed dataset. Specifically, we randomly selected 10,000 users and their 3,239,534 interacted micro-videos.
    \item \textbf{AmazonBooks} is a subset of the Amazon dataset~\cite{10.1145/2872427.2883037} too. We follow the HLLM~\cite{chen2024hllmenhancingsequentialrecommendations} to preprocess the dataset. Specifically, we retain only items and the users with at least 5 presences. We only keep item\_id ignore other features. The maximum length of context is set to 50. AmazonBooks contains 1,689,188 samples, 694,898 users, 10,053,086 interactions and 686,624 goods.
    % \item \textbf{Production} is an one-week period dataset with several sequential features, sampled an collected from Huawei Browser video feed recommendation. We use the most recent 1-day samples as the test set and the previous 6 days of data as the training set.
    % , such as doc\_id, category\_id, short-term interest topic\_id, and anonymous data masking user\_id.
\end{itemize}

\subsubsection{Metrics} We evaluate performance using gAUC~\cite{10.1145/3219819.3219823}, AUC~\cite{guo17deepfm} and Logloss. \textbf{AUC} measures the probability of a positive sample being ranked higher than a randomly chosen negative one. \textbf{gAUC} is to compute the AUC for each user separately, and then make a weighted average for them. In the experiment with large scale model, the metrics are \textbf{Recall@K (R@K)} and \textbf{NDCG@K (N@K)}. To account for variability, each experiment is repeated 5 times, and we report the average results.

\subsubsection{Implementation Details} The embedding size of all features are set to 64, and we concatenate item ID and category ID embedding to generate item embedding, then we concatenate other features and attention outputs as the input of MLPs to get prediction. Thus, the dimension of item embedding is 128. The maximum length of context is set as 100. The dimension of position embedding size of CAPE is grid searched in [16, 32, 64]. For multi-head attention models, the head number is grid searched in [1,2,4, 8]. We optimize all SR models with Adam optimizer with learning rate 5e-4 and 1e-3 in AmazonElectronics and KuaiVideo respectively. The batch size is set as 1024 in AmazonElectronics and 4096 in KuaiVideo. We adopt early stop with the patience of 3 epochs.

\subsection{Evaluation in Public Benchmark Datasets}

We first report results in the public benchmark datasets AmazonElectronics and KuaiVideo, which are shown in Table~\ref{tab:main_result}. We report the result with best AUC, and all results are averages over 5 times training with different random seeds. It should be noted that AUC and gAUC improvement on the third decimal place can be considered as significant improvement for offline prediction. 

From Table~\ref{tab:main_result}, we can draw several insightful conclusions: 
\begin{enumerate}
    \item CAPE significantly and consistently improves gAUC and AUC with all backbones, and achieves SOTA performance. Na\"ive PE and CoPE deteriorate baselines at some circumstance, such as DMIN in AmazonElectronics and SASRec in KuaiVideo. RoPE improves all metrics too, but it unable to outperform other baselines consistently. Above results demonstrate the superiority of CAPE and necessity for SR specific PE method.
    \item CAPE achieves SOTA performance in both datasets. Meanwhile, na\"ive PE performs better than CoPE in KuaiVideo averagely, and RoPE outperforms CoPE averagely in both datasets. These results demonstrate the validity of assigning position of context items by using dissimilarity rather than similarity that in CoPE.
    \item Na\"ive PE performs better with target attention than self-attention, and CoPE performs better with self-attention. Because of the fusion method, CAPE achieves SOTA performance with both target attention and self-attention with smaller position embedding dimension, indicating the better flexibility of CAPE compare to baselines.
    % Meanwhile, na\"ive PE perform better than CoPE in KuaiVideo and worse in AmazonElectronics, and CAPE can achieve SOTA performance in both datasets. These results indicate the better flexibility of CAPE than baselines, owing to the gate-based fusion method. 
    % \item Compared to CoPE, 
    % Meanwhile, na\"ive PE perform better than CoPE in KuaiVideo and worse in AmazonElectronics, and CAPE can achieve SOTA performance in both datasets.
    
    % the average relative improvement of CAPE are   at gAUC and at AUC in AmazonElectronics respectively, meanwhile at gAUC and AUC in KuaiVideo. These results demonstrate the superiority of choosing dissimilarity rather than similarity like CoPE
\end{enumerate}

% pe outperform none, cape is the best, naive better than cope
% amazon cope占有，kuaivideo上naive占优，因为

\subsection{Evaluation with Large-scale Models}
In this experiment, we aim to evaluate the performance of CAPE as the model parameters increase, which are shown in Table~\ref{tab:main_result2}. Experiments were conducted solely with SASRec, as it is the only backbone model whose performance improves with increasing parameters in our experiment. We conducted the experiment in AmazonBooks dataset based on the source code provided by HLLM \footnote{\url{https://github.com/bytedance/HLLM}}. We only user item\_id feature and embedding size is 512. All Transformer blocks in SASRec have 8 head.

\begin{table*}[htp]
    \centering
    \caption{Evaluation on Large scale model with SASRec in AmazonBooks dataset. Recall and NDCG values are averages over 5 random seeds and only keep 5 decimal. The best results are bold, and the underline indicates the best within the same backbone.}
    % \vspace{-1mm}
    \resizebox{0.98\textwidth}{!}
    {\begin{tabular}{c|c|ccccc|ccccc}
    \toprule
         \# Blocks&PE&R@5&R@10&R@50&R@200&Avg Impv.(\%)&N@5&N@10&N@50&N@200&Avg Impv.(\%)\\\hline
         \multirow{3}{*}{2}&Na\"ive&5.46138&7.04363&12.36615&\underline{19.79934}&0.0&4.12205&4.63225&5.79039&\underline{6.90217}&0.0\\
         &RoPE&5.30841&6.82317&12.00351&19.28775&-2.86&3.99734&4.48555&5.61007&6.69936&-3.06\\
         &CAPE&\underline{5.53449}&\underline{7.06162}&\underline{12.46354}&19.51095&+0.23&\underline{4.18327}&\underline{4.67609}&\underline{5.80960}&6.89349&+0.66\\\midrule
         \multirow{3}{*}{8}&Na\"ive&5.31316&7.03946&12.99099&21.03707&0.0&3.93586&4.49230&5.78699&6.99132&0.0\\
         &RoPE&5.42857&7.08019&12.71685&20.43641&-0.06&4.05921&4.59212&5.81769&6.97307&+1.41\\
         &CAPE&\underline{5.44455}&\underline{7.11573}&\underline{13.82449}&\underline{21.56204}&+3.12&\underline{4.06106}&\underline{4.59985}&\underline{5.84139}&\underline{6.99945}&+1.66\\\midrule
         \multirow{3}{*}{12}&Na\"ive&5.10464&6.77726&12.72448&20.30675&0.0&3.77088&4.30938&5.60076&6.82750&0.0\\
         &RoPE&5.43073&7.07688&12.75614&20.53542&+3.05&4.04065&4.57147&5.80790&6.97260&+4.76\\
         &CAPE&\underline{\textbf{5.50326}}&\underline{\textbf{7.18034}}&\underline{12.89601}&\underline{20.70926}&+4.27&\underline{\textbf{4.10001}}&\underline{\textbf{4.63976}}&\underline{5.88204}&\underline{7.05137}&+6.17\\\midrule
         \multirow{3}{*}{16}&Na\"ive&4.96879&6.67969&12.58460&20.68738&0.0&3.63913&4.18976&5.47254&6.68626&0.0\\
         &RoPE&5.41332&7.08177&12.82435&20.65328&+4.18&4.01359&4.55058&5.79979&6.97170&+7.29\\
         &CAPE&\underline{5.43980}&\underline{7.13070}&\underline{\textbf{12.92321}}&\underline{\textbf{20.77545}}&+4.84&\underline{4.05528}&\underline{4.60047}&\underline{\textbf{5.89149}}&\underline{\textbf{7.06736}}&+8.65\\
         
    \bottomrule
    \end{tabular}}
    % \vspace{-1mm}
    \label{tab:main_result2}
\end{table*}

From Table~\ref{tab:main_result2}, we can draw following insightful conclusions:
\begin{enumerate}
    \item In both large and small models, CAPE outperforms the baseline across almost all metrics, with an average improvement of up to 8.65\% in NDCG and 4.84\% in Recall, respectively. When the model has only 2 blocks, RoPE performs worse than Na\"ive PE. Even when the number of blocks increases to 8, the performance of the two baselines remains comparable.This result demonstrate the effectiveness of CAPE with large scale SR models. 
    \item These results demonstrate the excellent flexibility of CAPE, which enhances the performance of both large and small models. Na\"ive PE is only effective in shallow models, while RoPE requires large to a certain model depth to surpass Na\"ive PE (\# Blocks > 8). As the model becomes deeper, Na\"ive PE begins to have a negative impact, as models using Na\"ive PE experience a decline in average performance with increasing depth. In contrast, models using CAPE continue to improve. 
    \item CAPE also enhances model scalability. When the number of model blocks exceeds 8, the average performance of models using Na\"ive PE begins to decline. When the number of layers exceeds 12, models using RoPE also start to show a decrease in average performance. However, models using CAPE continue to improve in average performance even when the number of layers reaches 16. This result indicates the significant potential of CAPE for future large-scale recommendation models.
\end{enumerate}

\subsection{Ablation Study}
We conduct ablation study on the AmazonElectronics dataset, as its average context length is significantly longer than that of the KuaiVideo. This allows us to examine the impact of a longer context. We choose DIN (a classic target attention model) and SASRec (a classic self-attention model) as backbones. The evaluation metric in following experiments is AUC. %The position embedding dimension of both Na\"ive PE and CoPE is set same as item embedding, \textit{i.e.}, 128.

\subsubsection{Context length and position embedding dimension}

\begin{figure}[htp]
    \centering
    % \subfigure{\includegraphics[width=0.25\textwidth,height=0.25\textwidth]{fig/mgda.png}}\hspace{-2mm}
    \subfigure[DIN]{\includegraphics[width=0.45\textwidth,height=0.25\textwidth]{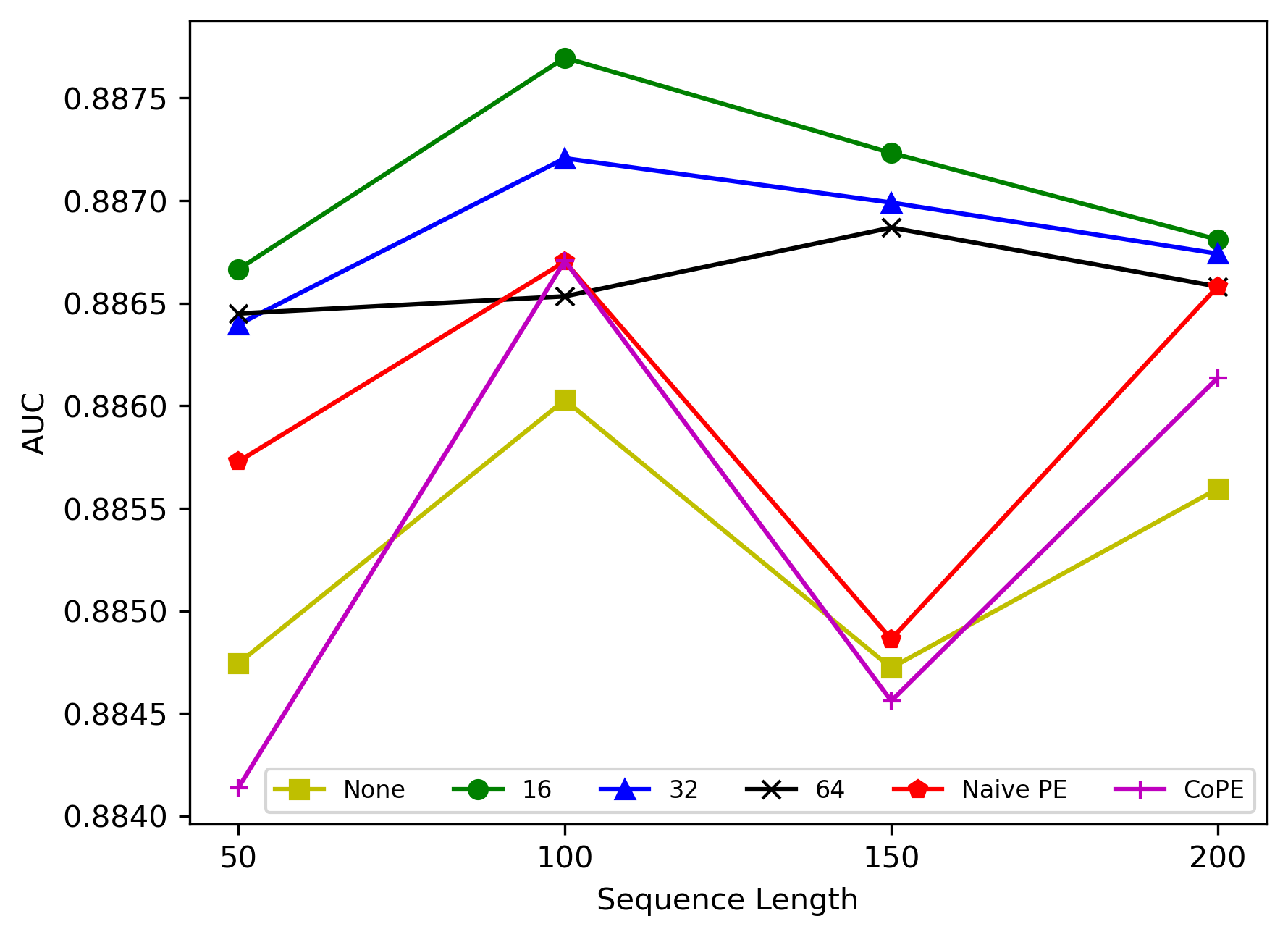}}
    \subfigure[SASRec]{\includegraphics[width=0.45\textwidth,height=0.25\textwidth]{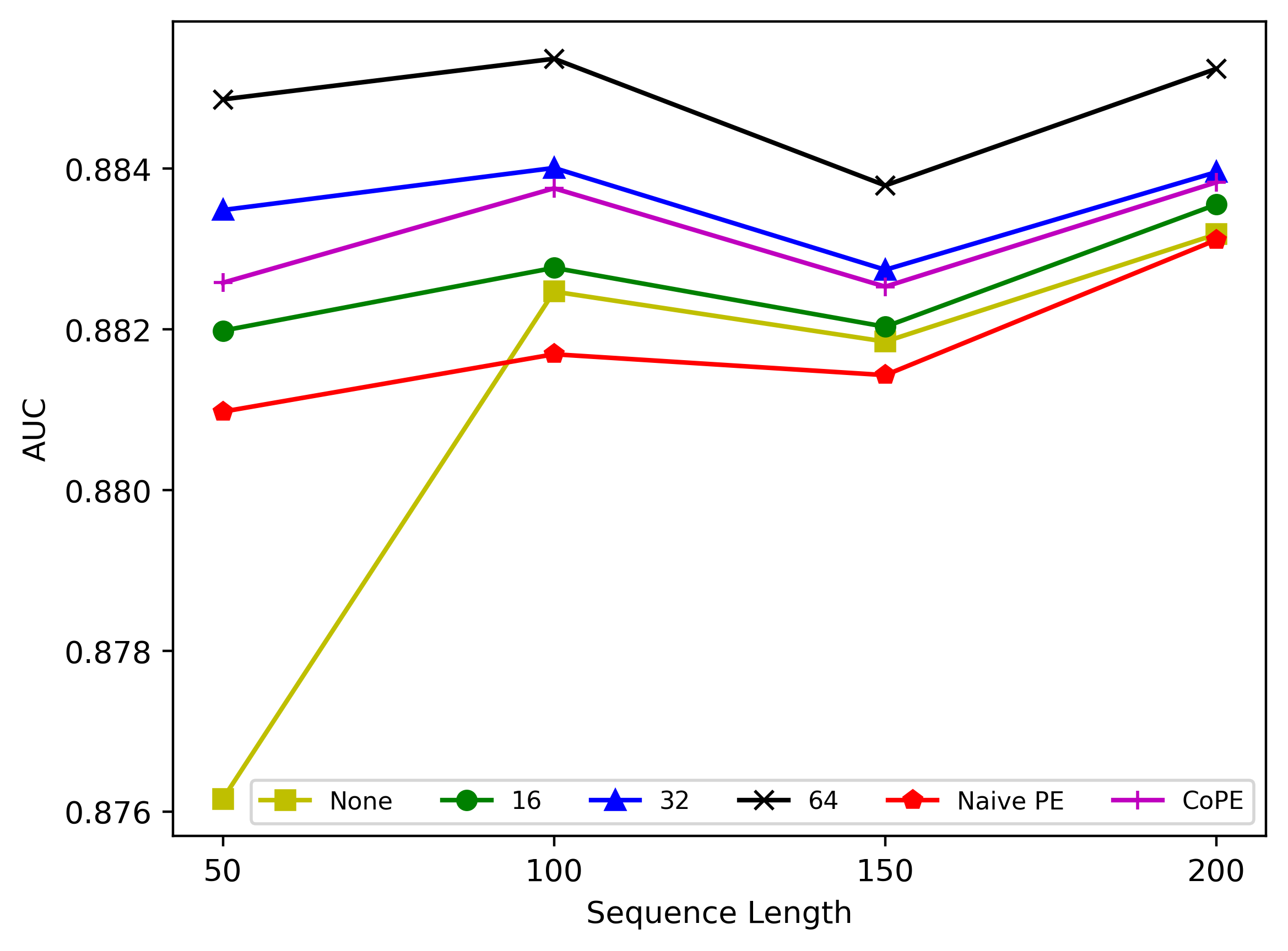}}
    \caption{Ablation study on context length and position embedding dimension in AmazonElectronics. When position embedding dimension is 32, CAPE is able to outperform all baselines across all context length.}\label{fig:ablation1}
    \vspace{-5mm}
\end{figure}

In this experiment, we aim to evaluate the impact of context length and position embedding dimension on CAPE. The experimental results are presented in Figure~\ref{fig:ablation1}. CAPE can set different position embedding dimension to not only save training and memory cost, but also achieves SOTA performance across all context lengths. This result further highlights the necessity of designing specific position encoding methods for sequential recommendation. 

% dimenion的作用

From Figure~\ref{fig:ablation1}, we can draw following insightful conclusions: 

1) With the DIN backbone, when the position dimension is less than or equal to 32, CAPE outperforms all baselines. Additionally, Na\"ive PE outperforms CoPE when the context length is either smaller or larger than 100. These results highlight the effectiveness of our fusion method. Since DIN attention differs significantly from the dot product attention used in NLP, our fusion method is able to handle this difference, whereas CoPE is not. 2) Although CoPE outperforms None and Na\"ive PE with SASRec, CAPE surpasses CoPE when the position embedding dimension is greater than or equal to 32. This result emphasizes the rationale behind using dissimilarity to assign positions in SR, as opposed to similarity-based assigning methods like CoPE. 3) CAPE enhances two SR models across different context lengths, while baseline PE methods significantly degrade SR models performance when context length is smaller than 100. This result demonstrates the better ability of CAPE to handle shorter user history compared to baselines.

\subsubsection{Context length and fusion methods}

As discussed in Section~\ref{method}, the goal of this experiment is to evaluate the impact of different fusion methods on CAPE. "none" in the figure means no position encoding method. The naive fusion method in this experiment simply adds the position embedding to the item embedding, so the position embedding dimension of naive fusion is same as item embedding. While the dimension of gate fusion method is set to best configuration in aforementioned experiment, \emph{i.e.} 16 with DIN and 64 with SASRec.

\begin{figure}[htp]
    \centering
    % \subfigure{\includegraphics[width=0.25\textwidth,height=0.25\textwidth]{fig/mgda.png}}\hspace{-2mm}
    \subfigure[DIN]{\includegraphics[width=0.45\textwidth,height=0.25\textwidth]{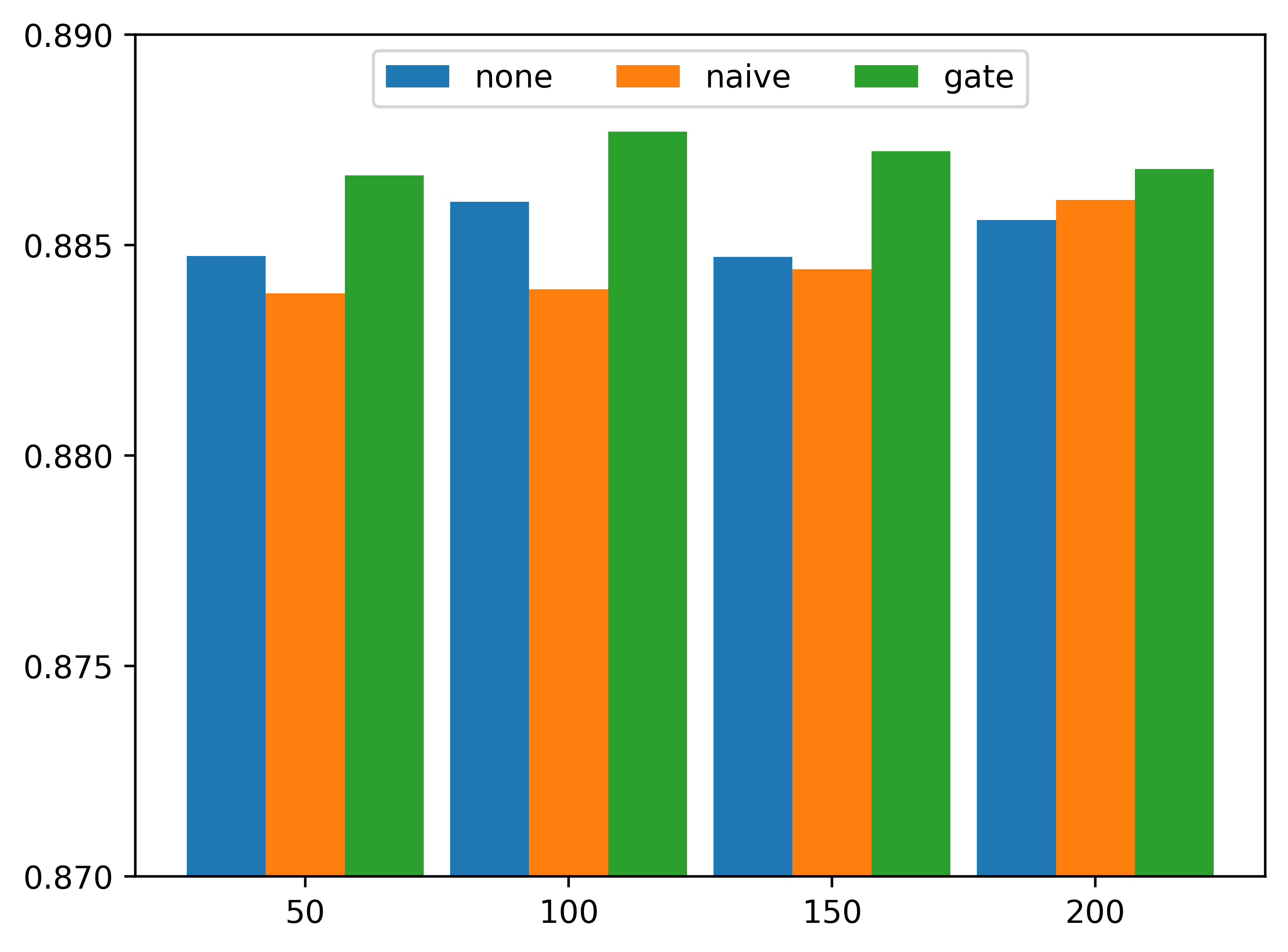}}
    \subfigure[SASRec]{\includegraphics[width=0.45\textwidth,height=0.25\textwidth]{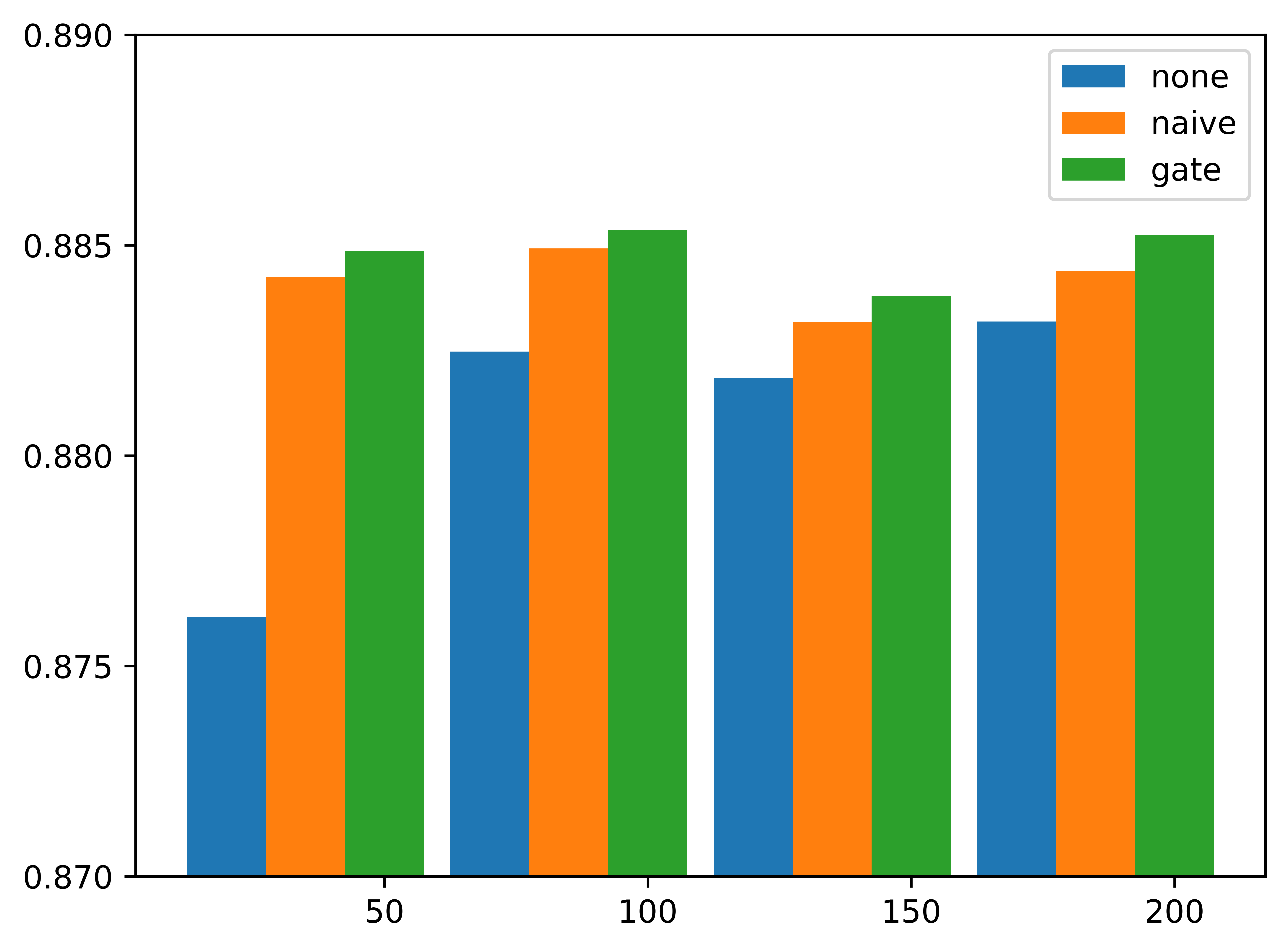}}
    \caption{Comparison of different fusion methods in AmazonElectronics. Naive fusion method deteriorates DIN, and None method deteriorates SASRec. CAPE enhances two attention type models significantly.}\label{fig:ablation2}
    \vspace{-1mm}
\end{figure}

The experimental results are presented in Figure~\ref{fig:ablation2}. From the figure, we observe that gate fusion method improves both DIN and SASRec across different context lengths. However, naive fusion method enhances SASRec but deteriorates the performance of DIN significantly. This result indicates that position embedding and item embedding may not lie in same hidden space when choosing DIN attention. And CAPE can boost both DIN attention and dot product attention simultaneously. Above results demonstrate the validity of introducing gate architecture of CAPE.

\subsection{Online A/B Testing}
The primary objective of the online experiments is to thoroughly evaluate the effectiveness of our proposed CAPE in a real-world applications. Specifically, the experiments were conducted within the most valuable channel of Huawei AppGallery. The integration of CAPE into the existing ranking model is seamless, and the process required minimal engineering effort, as no significant modifications were needed within the online serving infrastructure. We conducted a comprehensive 8-day online A/B test, and the results are shown in Table~\ref{tab:online}, provide strong evidence of CAPE's effectiveness.

From the table, it is evident that our method consistently outperforms the baseline, resulting in a daily improvement of the north star metric, \emph{eCPM} (Effective Cost Per Mille). This daily improvement in \emph{eCPM} underscores the reliability and consistency of CAPE’s impact on performance. During the course of the experiment, CAPE demonstrates an average improvement of 3.62\% relative to the highly optimized online baseline, marking a significant enhancement in the system’s overall efficiency and effectiveness. These results validate the efficacy of our context-aware position encoding mechanism, showcasing its potential to deliver tangible improvements in the online recommendation system. 

\begin{table}[h]
\caption{Online results of a eight-day online A/B testing. The north star metric is eCPM.}
% \vspace{-3mm}
\resizebox{0.45\textwidth}{!}{
\begin{tabular}{cccc|c}
\toprule
Day 1 & Day 2 & Day 3 & Day 4  &\multirow{2}{*}{\textbf{Average} } \\
\cmidrule{1-4}
+6.12\% & +4.85\% & +2.12\% & +0.19\% \\
\cmidrule{1-4}
Day 5 & Day 6 & Day 7 & Day 8 &\multirow{2}{*}{\textbf{+3.62\%} } \\
\cmidrule{1-4}
+3.06\% & +3.08\% & +0.80\% & +4.49\%  \\
\bottomrule
\end{tabular}
}
% \vspace{-3mm}
\label{tab:online}
\end{table}

% App Gallery offline
% \begin{table*}[htp]
% \centering
%     \begin{tabular}{c|cccccccc}
%     \toprule
%     method&Task 1&Task 2&Task 3&Task 4&Task 5&Task 6&Task 7& Task 8\\\hline
%     baseline&0.917927&0.891101&0.923967&0.935712&0.928801&0.894898&0.910286&0.748306\\
%     baseline+CAPE&\textbf{0.91949}&\textbf{0.892462}&\textbf{0.927646}&\textbf{0.936301}&\textbf{0.930169}&\textbf{0.895593}&\textbf{0.911068}&\textbf{0.749173}\\
%     \bottomrule
%     \end{tabular}
%     \caption{Offline experiment in industrial dataset. The metric is AUC and only keep 6 decimal. The best results are bold, and CAPE significantly boost baseline at every task.}
% \end{table*}

\section{Conclusion and Future work}
In this paper, we propose CAPE, a Context-Aware Position Encoding for sequential recommendation. To the best of our knowledge, CAPE is the first PE method specifically designed for sequential recommendation. CAPE is a context-dependent position encoding method that represents various levels of positional abstraction within the user context. Extensive experiments on public benchmark sequential recommendation datasets, as well as online A/B testing on real-world commercial platform, demonstrate the effectiveness of our approach. Leveraging the characteristics of sequential recommendation, we employ dissimilarity along with a gate architecture to assign position to each items in context. CAPE consistently improves five mainstream SR backbones in two benchmark datasets, whereas conventional PE methods do not. Not only that, CAPE can enhance both effectiveness and scalability of large-scale recommendation model, indicating the potential of CAPE for future large-scale SR models. 

Our research highlights the necessity of developing specialized position encoding techniques for sequential recommendation, which should attract the attention of the research community. In the future, we aim to evaluate the effectiveness of CAPE across a wide range of recommendation tasks, such as multi-domain and multi-task sequential recommendation. Lastly, we will conduct more exploration in theory to deeply understand the nature of position information in sequential recommendation.

\clearpage
%%
%% The next two lines define the bibliography style to be used, and
%% the bibliography file.
% \clearpage
\bibliographystyle{ACM-Reference-Format}
\balance
\bibliography{sample-base}

%%
%% If your work has an appendix, this is the place to put it.
% \appendix

% \section{Research Methods}

\end{document}